\documentclass[vecphys]{svmult}

\usepackage{makeidx}         
\usepackage{graphicx}        
\usepackage{multicol}        
\usepackage[bottom]{footmisc}
\usepackage{natbib}
\usepackage{journals}

\makeindex             

\begin{document}

\title*{From Multiwavelength to Mass Scaling: Accretion and Ejection
  in Microquasars and AGN}

\titlerunning{From Multiwavelength to Mass Scaling}
\author{Sera Markoff}

\institute{Astronomical Institute ``Anton Pannekoek'', University of
  Amsterdam, Kruislaan 403, 1098SJ Amsterdam, the Netherlands
\texttt{s.b.markoff@uva.nl}
}

\maketitle
\abstract{A solid theoretical understanding of how inflowing, accreting plasma around black holes and other compact objects gives rise to outflowing winds and jets is still lacking, despite decades of observations.  
The fact that similar processes and morphologies are observed in both X-ray binaries as well as active galactic nuclei has led to suggestions that the underlying physics could scale with black hole mass, which could provide a new handle on the problem.  In the last decade, simultaneous broadband campaigns of the fast-varying X-ray binaries particularly in their microquasar state have driven the development of, and in some cases altered, our ideas about the inflow/outflow connection in accreting black holes.  Specifically the discovery of correlations between the radio, infrared and X-ray bands has revealed a remarkable connectivity between the various emission regions, and argued for a more holistic approach to tackling questions about accretion.  This article reviews the recent major observational and theoretical advances that focus specifically on the relation between the two ``sides'' of the accretion process in black holes, with an emphasis on how new tools can be derived for comparisons across the mass scale.  }

\section{Introduction}\label{sec:1}

The process of disentangling inflow from outflow in accreting black holes is something of an exercise in semantics as we approach the event horizon. What we have traditionally thought of as ``inflow'' solutions, such as the various flavors of radiatively inefficient accretion flows (RIAFs; e.g. \cite{NarayanYi1994}), are now thought to
become unbound, resulting in windy ``outflows'' from their surfaces, without even taking jet formation into consideration (e.g., \cite{BlandfordBegelman1999}).  Thus the problem is far from simple, and the jets which are somehow eventually launched could tap some combination of bound flow, unbound flow and black hole spin energy for plasma and power.  Until we understand which class of solution dominates, correctly interpreting the emission from the innermost regions of the accretion flow(s) will remain a challenge.  Yet this region is precisely where the most extreme physics is lurking, and thus what interests many scientists in the field of astrophysical accretion rather acutely.  

From the outermost reaches of the jets to the inner boundary of the accretion flow, the ensuing radiation spans the entire broadband spectrum, including the lowest frequency radio waves through TeV $\gamma$-rays.  Such a range, of almost 20 decades in frequency, can provide an enormous lever arm for theoretical modeling if the accreting system is viewed holistically.  As I will discuss further below, the lever arm supplied by multiwavelength data is likely our best hope for actually breaking the theoretical degeneracy currently hindering our progress.  However, because we are not yet at the point of achieving fully self-consistent a priori physical solutions (though that is certainly on the horizon, see article in Chap.~9), we must first make a set of assumptions relating the two ``sides'' of accretion before we can build models to test those assumptions against the data.  Eventually such approaches can provide the groundwork and boundary conditions for more exact numerical methods.

This chapter will in some ways take over where the last left off, with an emphasis on the recent development and evolution of our theoretical ideas specifically in the context of the increasing prevalence of multiwavelength campaigns.  I will focus on documenting the process by which our ideas have undergone shifts, based on this new approach to observations, and the far-reaching consequences that black hole binaries (BHBs) may now have for deconstructing galaxy evolution on cosmological timescales, assuming mass-scaling is really all it is cut out to be.   My obvious bias is towards understanding jets in these systems, but I try to touch on all scenarios and give references for further reading whenever possible.

\section{Changing paradigms}\label{sec:paradigms}

Up until surprisingly recently, the standard view of BHBs did not
include jets, even though one of the most quoted papers in the field
of accretion shows a nice figure of an outflow! --(I am referring to
Fig. 9 in \cite{ShakuraSunyaev1973}, which shows an X-ray emitting
``cone'').  For example, in the now canonical textbook \emph{X-ray
  Binaries} released well after radio jets were a commonly known
feature of BHBs \cite{Lewinetal1995}, only one article discusses
them at all, and then only as a separate feature not integrated into
the basic picture.  Historically speaking, this oversight is not so surprising.
When X-ray binaries (XRBs) were first discovered via sounding rockets
in the late 1960's, accretion was the obvious culprit and the idea of
viscously-dissipating flows quickly took over as the predominant
paradigm.  The discovery of variable, hard X-ray emission necessitated
the invention of a region called the corona (see, e.g.,
\cite{EardleyLightmanShapiro1975,ShapiroLightmanEardley1976}), where
hot electrons reside and upscatter thermal disk photons into a hard
power-law tail.  The concept was eventually developed into a fairly
self-consistent picture by \cite{HaardtMaraschi1991}.  With many
varieties now spanning the parameter space of configurations including
reflection (e.g., \cite{SunyaevTitarchuk1980,Titarchuk1994,TitarchukLyubarskij1995,Magdziarz1995,PoutanenSvensson1996,Coppi1999}), this idea of a region of hot plasma in stasis above or within the cooler accretion flow, radiating relatively isotropically, has been the canonical viewpoint for decades, and remains so for many researchers today.
 
The problem, as I see it anyway, is that these models do not take into account the role of strong magnetic fields in the inner regions of the accretion flow.  At the time of their development their omission was understandable, because the study of magnetic phenomena in astrophysics was significantly less developed than it is today.  However in the intervening years we have strong evidence that magnetic fields can be generated in accretion flows via the magneto-rotational instability (MRI), as well as brought in from the environment via the accreting gas (see more about magnetohydrodynamics, or MHD,  in Chap.~9).  The inclusion of magnetic fields will not only change many of the underlying assumptions in the classical thermal Comptonization picture, but is also necessary in order to begin to address the relationship between the inflow and the obviously magnetic phenomena of jets.  

Interestingly, the necessity of strong magnetic fields in various BHBs was realized decades ago by \cite{Fabianetal1982}, who proposed that the fast optical flaring observed in the BHB GX~339-4 was the result of cyclotron emission, as mentioned also in Chap.~9.  Building on this approach, more recent works have considered these so-called magnetic coronae \cite{diMatteoCelottiFabian1997,WardzinskiZdziarski2000,MerloniFabian2002,PoutanenVurm2008}, all of which involve similar distributions of hot electrons as the static corona models, but which in addition also include magnetic effects.  It is hard to imagine a world in which such a plasma has no relationship to the hot, magnetized plasma required at larger scales to power the nonthermal synchrotron emission observed in BHBs in the hard and intermediate states.  One approach suggested by \cite{Beloborodov1999} and later \cite{MalzacBeloborodovPoutanen2001} is that radiation pressure drives the magnetic flares away from the disk at mildly relativistic speeds, a scenario that begins to sound rather like the base of a jet. However this idea was motivated in part by the need to reduce the problematically high reflection fraction resulting from static disk-corona models \cite{Gierlinskietal1997}, because the mild beaming reduces the fraction and energy of disk photons available for Compton scattering in the jet frame.  The authors also did not consider a relationship between this beamed plasma to the outflows explicitly.  If one considers the presence of large scale ordered magnetic fields at the inner edge of the accretion flow, as will likely result if any strong field is brought in or generated, then once the accreting plasma is ionized, it will more or less follow the field line configurations and the idea of a static corona is fairly unlikely.   

By now it is generally recognized that there must be some kind of relationship between the corona and the jet outflows, as evidenced by the tight radio/infrared(IR)/X-ray correlations discussed in more detail in other chapters of this volume.  The exact nature of this connection is where the uncertainty creeps in. Is it simply a matter of reservoirs of magnetic energy being tapped by both the corona and jet respectively, as suggested by, e.g., \cite{MalzacMerloniFabian2004}, or does a magnetized corona actually directly feed the jets as my collaborators and I have explored \cite{MarkoffNowak2004,MarkoffNowakWilms2005}?  The difficulty in determining just how absolute the inner disk/coronal outflow/jet connection is (i.e., are they both feeding from the same trough or does the jet feed off of the corona alone) can be attributed to the problems we currently have in 
disentangling the associated spectral contributions and timing signatures of the various components.  

I think all of us writing in this volume believe that the quality of the data currently available already has the potential to address these gaps in our understanding, if only we could figure out the right questions to ask!  The combination of timing (see Chaps. 3 \& 8) and spectral (see other Chapters) features is providing many clues, and one of the primary goals of this article is to summarize the recent history of how these observations have defined and, in some recent cases, altered our ideas about the accretion/outflow connection in accreting BHs. 
As I discuss the process of discovery itself, I emphasize that while collecting phenomenology is useful, the major steps forward did require having some theoretical frameworks already in place.  We should keep this fact in mind as we consider our future lines of attack.  

\section{The driving observations and some interpretation}\label{sec:obs}

As described in other Chaps. of this volume, there are two types of jets observed in BHBs:  the compact, steady, and self-absorbed flows associated with the low-hard state (LHS) and at least the brighter quiescent states, and the optically thin, more discrete ejecta associated with the transition from the hard to soft intermediate states (HIMS to SIMS), near what we call the ``jet line" \index{jet line}(see Chaps. 3 and 5).   The steadiness of the LHS jets, as well as their compactness, makes them much easier to model since we are essentially observing their outer layers, on a size scale selected out by the stratification of frequency-dependent synchrotron self-absorption (see Chap. 7).  Therefore in this state we can treat variability, at least to first order, as due to changes in the input power to a steady flow where the same essential geometry persists at all times.  Such an assumption, if merited, simplifies the physics significantly.  Secondly, we can justify ignoring radial stratification in the jets, because while it is likely an indisputable fact of jet evolution (Chaps.~7 \& 9), to first order we are only able to observe the outermost layers that photons can escape, though for timing analysis stratification can no longer be neglected.  For SED modeling, however, these two factors explain why the LHS is the favorite target of theorists so far, and indeed despite our ignorance about many important factors of jet internal physics, we are still able to make headway, and some sensible predictions.  

In this section I will summarize briefly how jets have been incorporated into the standard BHB picture, to a great extent based on earlier work modeling active galactic nuclei (AGN).   I will then complement the discussion of scaling in Chap.~5 to include some basic jet physics, and how it has been used to interpret the radio/IR/X-ray correlations, and led directly to the discovery of the Fundamental Plane of black hole accretion (\cite{MerloniHeinzDiMatteo2003,FalckeKoerdingMarkoff2004}, and see Chap.~5).   Finally I will describe where I think new work needs to be focused, and along the way touch on other salient points regarding the inflow/outflow connection, with a distinct outflow-biased perspective.

\subsection{Multiwavelength correlations}\label{subsec:corrs}

The Galactic source GX~339-4\index{GX~339-4} has been one of the most important testbeds for the development of our ideas about accretion/ejection in BHBs.  After the first detection of correlated optical/X-ray flaring \cite{Motchetal1982}, the observational community became more interested in looking for simultaneous variability at multiple wavelengths.  The results were extremely fruitful, starting with a campaign by \cite{Hannikainenetal1998} using the Molonglo Observatory Synthesis Telescope (MOST) to monitor GX~339-4 in the radio over the course of several years.  The authors reported initial evidence for a correlation between the radio flux and the X-ray fluxes from both the \emph{RXTE} and \emph{BATSE} orbital observatories.  The log-linear correlation was later confirmed to appear in the LHS, with the exact power-law index constrained by, e.g.,  \cite{Corbeletal2000,Corbeletal2003}.  Further developments are discussed in Chaps.~4 \& 5, but most recently it has been found that the radio/X-ray correlation is also echoed by correlations between the X-rays and the near-infrared (NIR) \cite{Russelletal2006}, and also shows deviations from a single power-law.  Regardless of the exact slope of the relation, the correlations between the radio, NIR and X-rays belie an intimate connection between the inner accretion flow and the jets, and has been the driver behind much of the recent development in our understanding of accretion physics.  

In retrospect, some form of correlation should not have been very surprising;  matter falls in, replete with angular momentum, and is somehow subsequently expelled, or some portion of its momentum serves to spin up the black hole which then provides energy for jet-powering Blandford-Znajek \cite{BlandfordZnajek1977}\index{Blandford-Znajek mechanism} processes.  In either scenario, the two processes of inflow and outflow are clearly related.  What is more surprising is that, even within the observed deviations, there is a relatively stringent relationship that holds in as many sources for which we can obtain good simultaneous data.  As soon as this bond was hinted at by work in \cite{GalloFenderPooley2003}, the radio/X-ray correlation in particular presented itself as a chance to explore the physics driving accretion/ejection in BHBs as a class, and a new way to probe the conditions responsible for jet launching.

As discussed elsewhere in this volume, we used to think that the slope of the radio/X-ray correlation was fixed at $L_R \propto L_X^m$, where $m\sim 0.7$, though now we see evidence for scatter (even for the same source) around that value, in a range of more like $0.6-0.8$.  Even so, such a small range adhered to by all sources with simultaneous radio and X-ray (or IR) data is suggestive of something universal, and it turned out to be an important baseline in extending our understanding of jets to incorporate potential scaling of the underlying physics with mass.   A predictable scaling with mass for the physics of accretion around black holes would automatically elevate BHBs from tiny, ISM-churning contributors to the X-ray background (in the eyes of the AGN community) to tiny, but very interesting replicas of AGN with smaller dimensions and faster timescales.  Given the importance of AGN cycles for galaxy growth and cluster evolution, it is no wonder that the hint of such usefulness has recently raised the profile of BHB studies with researchers working on these problems.  

In the following sections I will review how we established the first concrete mapping of a BHB accretion state to at least one AGN class, and how this sequence can serve as a model for any potential further mappings for the more complex, powerful and more transient accretion states.

\subsection{Accretion/ejection correlations, simple models and scalings}\label{subsec:jetspec}

\begin{figure*}
\centerline{\includegraphics[width=\textwidth]{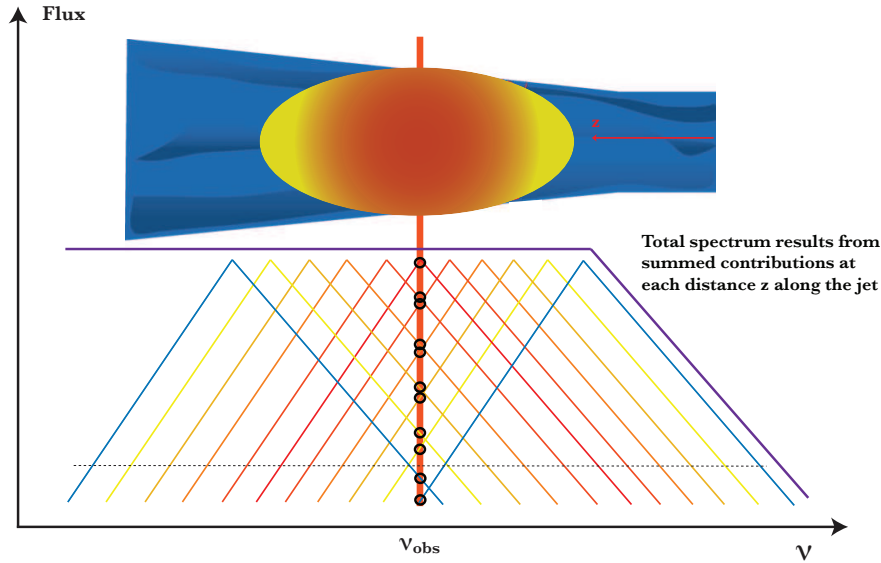}}
\caption{Schematic illustrating the stratified spectrum and
  frequency-dependent photosphere of an idealized, self-absorbed
  synchrotron jet. Each segment of the jet contributes approximately
  the same peaked, self-absorbed spectrum that combine to give the
  ``hallmark'' flat total spectrum.  A telescope observing at
  $\nu_{\rm obs}$ will see the largest contribution from the segment
  whose individual spectrum peaks at that frequency, and increasingly
  smaller contributions from neighboring segments (fluxes of each
  segment at $\nu_{\rm obs}$ indicated by black circles), generally
  producing an elongated Gaussian ellipse-like photosphere.  The
  visible jet at a particular frequency is thus much smaller in scale
  than the actual jet in the case of high optical depth.  Figure clearer in color (electronic version).}
\label{fig:jet}
\end{figure*}

\textbf{\textit{The compact jet paradigm}}
\newline
The compact jets observed in the LHS appear quite similar to those seen in the weakly accreting class of low-luminosity AGN (LLAGN; e.g. \cite{Ho1999}), or in the inner cores of extended AGN jets.  Their radio spectra are flat to inverted ($F_\nu\propto \nu^\alpha$, where $\alpha\sim0.0-0.3$), a spectral characteristic that together with high brightness temperatures and polarization are the ``hallmark'' signatures of compact jet synchrotron emission.  It is actually quite difficult to get such a spectrum from other scenarios, because self-absorbed synchrotron emission from a given distribution of leptons moving relativistically in a magnetic field results in a peaked spectrum, with a $\nu^{5/2}$ spectrum in the optically thick portion.   Similarly, beyond the optically thick-to-thin break one usually detects a declining power-law that reflects the underlying lepton distribution in the optically thin regime (see, e.g., \cite{RybickiLightman1979}). Thus a flat spectrum requires the radiating medium to be not only optically thick, but stratified; that is, the self-absorption frequency must vary with the spatial scale of its origin in the system.  Ever since the classical work by \cite{BlandfordKoenigl1979} (and see also Chap.~7) we have understood that the flat/inverted spectrum is a natural result of a steady jet where plasma properties are conserved, and energy partition between the magnetic field and hot and cold plasma is fixed, in a smoothly expanding flow.  These initial conditions provide the right ``conspiracy'' where each region of the jet contributes roughly the same spectral shape, with peak flux occurring lower in frequency the further out in the jet it originates (see Fig.~\ref{fig:jet}).  As a direct result, the photosphere changes in location on the jet as a function of observing frequency.  Such an effect has been empirically tested, and is known as core shift.  The extent of inversion in the radio slope depends on the exact scaling of density and magnetic field with distance from the launching point, the radiating lepton particle distribution, the cooling and reacceleration functions, and the jet dynamics.

Generally, imaging the jets responsible for detected flat/inverted radio spectra is challenging because of the optical depth.  At a given frequency, one cannot observe the entire jet but just the photosphere, which will look rather more like a Gaussian ellipse implying elongation beyond what is expected from, e.g., an accretion flow.  For instruments with very good sensitivity, a deep look at a flat spectrum source will often turn up a resolved jet. For instance a VLBI survey of over 100 AGN showing flat core features resulted in imaged jets in all brighter sources \cite{Kellermannetal1998}.  I therefore think it is a safe assumption that any source showing a flat/inverted radio/NIR spectrum with a high brightness temperature is very likely due to a compact jet. 

In LLAGN the radio spectra tend to peak, sometimes with an additional excess, or``bump'', in the submillimeter (submm) band.  If one wanted to naively begin considering any sort of mass scaling, one could imagine what a jet would look like if the same relative (in Eddington accretion units $\dot{m}_{\rm Edd}$) power were fed into the same relatively sized region (expressed in $r_g=2GM/c^2$), but for a 10$M_\odot$ black hole instead.  The most general prediction is that BHBs would exhibit a flat/inverted spectrum up to a break frequency somewhere well above the submm regime, reflecting the more compact scales and thus higher particle and magnetic energy densities for the same fraction of Eddington power.  It turns out that this prediction is indeed empirically confirmed, but it took some time before the necessary observations were in place.  

\begin{figure*}
\centerline{\includegraphics[width=\textwidth]{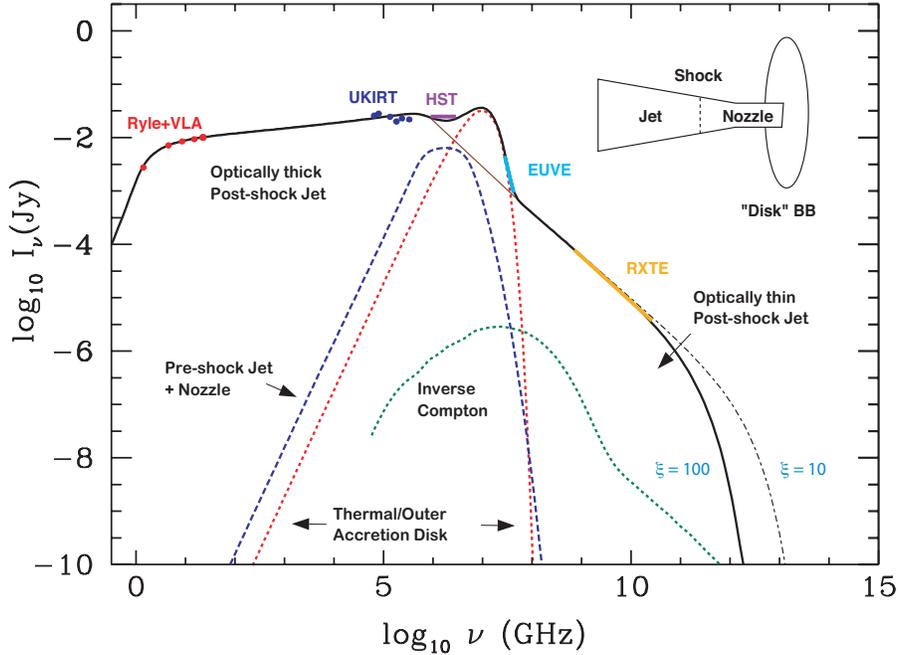}}
\caption{Application of a mass-scaling jet model to the first quasi-simultaneous multiwavelength data set of a BHB including UV data, for the Galactic source XTE~J1118+480.  The smaller scales compared to AGN results in a prediction of the jet synchrotron spectral break in the OIR band, and a significant contribution of jet synchrotron to the X-rays (see \cite{MarkoffFalckeFender2001} for details). }
\label{fig:j1118}
\end{figure*}

\vspace{.1in}\noindent\textbf{\textit{Flux correlations and mass scaling}}\newline
\index{empirical correlations}\index{scaling}
It is easy to take for granted that multiwavelength campaigns have always been the norm for BHB studies, as is the case for AGN blazars, but it was really only at the time that the radio/X-ray correlation was being determined that observers were in general awakening to the realization that simultaneous broad-band spectra of BHBs were not only interesting, but vital to capture the quick transitions observed in these sources.  In the year 2000, not even a decade before the writing of this volume, a new BHB was discovered that resided at high enough Galactic latitude to be visible to the \emph{EUVE} instrument, XTE~J118+480  \index{XTE~J118+480}\cite{Remillardetal2000}.  An extensive campaign was performed, really the first of its kind, involving quasi-simultaneous radio through submm \cite{Fenderetal2001} and IR through X-ray \cite{Hynesetal2000} observations.  It seemed like an opportune chance to test whether (and how well) a jet model developed originally for a LLAGN (in this case, Sgr~A*; \cite{FalckeMarkoff2000,Markoffetal2001})\index{Sgr~A*} could work to describe a BHB in the LHS, believed to be accreting at a comparably low accretion rate.  The original idea was to model only the radio through IR with a compact jet, since observed variations suggested that the flat/inverted spectrum continued at least up until this frequency (e.g., \cite{Fender2001a}).  After reworking the original model to include all size and power scales in mass-scaling units of $r_g$ and $L_{Edd}$ only, we replaced the supermassive black hole with a stellar remnant of around $10 M_\odot$ and easily found a solution that went through the radio through IR data, indeed even extending into the optical and possibly the UV depending on the compactness of the jet nozzle region  (\cite{MarkoffFalckeFender2001}, and see Fig.~\ref{fig:j1118}).

A potential high-energy contribution from jet synchrotron radiation was an interesting result, and because of simple scaling arguments that I will summarize below, not altogether surprising.  A further result that we did not originally anticipate was that the broad continuum features of almost the entire spectrum could be fit with jet synchrotron emission, after only one change in assumptions between this application and the model for Sgr A*.  That change was to assume that a significant fraction of the particles entering the jet in a quasi-thermal distribution, as assumed for Sgr A*, are subsequently accelerated into a power-law starting at a location around 100-1000 $r_g$ from the jet base.  In Sgr A*, the source is accreting at $\leq 10^{-9} \dot M_{\rm Edd}$ and particle acceleration in the jets must be extremely weak or absent during quiescence (flaring may be another story; \cite{Markoffetal2001}), as indicated by the stringent limits on any power-law component in the optically thin IR regime (e.g., \cite{MeliaFalcke2001}).  One of the main results of this paper was to demonstrate that the combination of empirical fact (jets accelerate particles into power-laws with energy index 2-2.4) with conservative assumptions about the acceleration rate (see, e.g., \cite{Jokipii1987}), taken together with calculated cooling rates from synchrotron and Compton processes, implies that it is very difficult to suppress some level of synchrotron emission extending into the X-ray band for BHBs, even for ``low-luminosity'' systems.  
For the case of XTE~J1118+480 we showed an extreme scenario, although not in conflict with the actual data set, which displays no indications of disk reflection or spectral cutoff.  There are good reasons, however, to think that in general there is also a Compton scattered emission component (see below).

\subsection{Interpreting the correlations}\label{subsec:interp}

Because the synchrotron jet model presents an obvious connection between emission over decades in frequency, it seemed worth exploring its
relevance for explaining the radio/X-ray correlation.  We decided to
try fitting the same data that had established the correlation, using
a variation of this mass-scaling jet model, and interestingly found\index{scaling}
that we could explain the correlation by only changing the power (as a
fraction of accretion energy) input into the jets
\cite{Markoffetal2003}.  The correlation suddenly presented itself as
a tool with which we could probe scalings of various emission
processes with power as well as mass.  

It turns out that the reason this class of model is so successful at explaining the correlations follows from some fairly straightforward radiative physics within the framework of self-absorbed jet models, as well as some accretion flows.  Given how complicated the physics internal to jets seems likely to be, with turbulent fields, stratified layers of different velocities, and distributed acceleration processes, it is somehow rather amazing that the spectra we observe fall into such a small range of slopes, remarkably close to the predictions of the idealized case of \cite{BlandfordKoenigl1979}.  It is reminiscent of the situation with thermal accretion disks:  we know that these should be turbulent, dissipative and complex inflows, yet they appear to follow the simple multicolor black body paradigm rather well.  Ironically, the phenomenon of high opacity allows us to understand global properties of compact jets, despite our lack of knowledge about their detailed internal properties.  

To understand how a synchrotron jet model reproduces the $m=\sim 0.7$
index of the radio/X-ray correlations, one can think first of an
idealized jet spectrum which has flux $F_\nu \propto \nu^{\alpha_R}$
up to the synchrotron self-absorption break frequency, $\nu_{\rm
  SSA}$, where the most compact radiating region of the jet
(containing a power law of leptons with distribution $N_e \propto
E_e^{-p}$) becomes optically thin and thus afterwards has a
$\nu^{-\alpha_X}$ spectrum, where $\alpha_X=(p-1)/2$ is the synchrotron
spectrum.  The total spectrum is thus a broken power-law (see
Fig.~\ref{fig:jet}) and if one knows how the optically thick flux and
the break frequency scale with jet power, $Q_j$, it then comes down to
algebra to derive the predicted correlation index $m\equiv d\log L_R/d\log L_X$.  In the original derivation of the dependence of radio luminosity on power for an idealized, compact, conical jet, \cite{BlandfordKoenigl1979} found $L_R ({\rm i.e.}\;F_{R,\nu}) \propto Q_j ^{17/12}$, and the same scaling was subsequently derived in later models \cite{FalckeBiermann1995,Markoffetal2003} because of similar underlying assumptions about the dependence of internal pressures with distance along the jet.  

A more generic treatment of the scaling can be found in
\cite{HeinzSunyaev2003}, who explore a broader range of assumptions
for the various scaling accretion models.  For instance, whether a jet is assumed to launch specifically from a classical ADAF \cite{NarayanYi1994}\index{ADAF} or any other scenario where the input power is assumed to be simply proportional to the disk luminosity, similar scalings are predicted.   The magnetic energy density $U_B$ and lepton energy density $U_e$ have a fixed partition ratio (the simplest case being the assumption of equipartition) proportional to $\dot m/M_{\rm BH}$, where $\dot m\equiv \dot M/M_{\rm Edd}$ and is thus dependent on black hole mass $M_{\rm BH}$.  By using these scalings, with all distances 
in addition scaled in terms of $r_g(M_{\rm BH})$, in the expressions for self-absorbed synchrotron flux from, e.g., \cite{RybickiLightman1979}, one can directly calculate the total flux at a given frequency from the entire jet:
\begin{equation}
F_\nu=\int_{r_g}^{\infty} dr R(r) S_\nu(r),
\end{equation}
where $R(r)$ is the jet diameter at $r$ and $S_\nu$ is the synchrotron source function.  For either case of an ADAF or $Q_j \propto M\dot m$, following Eq. (12b) in \cite{HeinzSunyaev2003}, one finds the dependence on $\dot m$:
\begin{equation}\label{eq:dfdmdot}
\frac{\partial \ln F_{\nu,R}}{\partial \ln \dot m} = \frac{2p + (p+6)\alpha_R + 13}{2(p+4)}.
\end{equation}
For the ``canonical'' accelerated particle index $p=2$ one finds $F_\nu \propto \dot m^{17/12 + 2\alpha_R/3}$ which for a perfectly flat self-absorbed spectrum reproduces exactly the original Blandford \& K\"onigl result.  

To derive the correlation index $m$, we need to derive the predicted ratio of radio to X-ray fluxes. For the case of a model such as that presented in Fig.~\ref{fig:j1118}, this index is just the ratio of the (log) fluxes above and below $\nu_{\rm SSA}$, because the X-rays are due to optically thin synchrotron alone.  However for an arbitrary
X-ray process we need to insert its dependence on $\dot m$ as in (\ref{eq:dfdmdot}):
\begin{equation}
m=\frac{\partial \ln F_{\nu,R}}{\partial \ln F_{\nu,X}}=\frac{\partial \ln F_{\nu,R}}{\partial \ln \dot m}\frac{\partial \ln \dot m}{\partial \ln F_{\nu, X}},
\end{equation}
which for $p=2$ and an ADAF or $Q_j\propto M \dot m$ case gives $m=(17+8\alpha_R)/21$ and reproduces the observed $m\sim 0.7$ for a flattish radio spectrum.  

Clearly the exact value of the correlation will vary slightly from source to source, because we expect variations in the local acceleration particle index $p$, and in the geometry and compactness effecting $\alpha_R$ and $\alpha_X$.  The power of this simplistic analysis is that even accounting for such scatter, certain accretion processes can be ruled out entirely already, based solely on the data.  For instance if the X-rays originate in a perfectly efficient corona where $F_\nu,X\propto \dot m$ then $m\sim 1.4$, which has never been observed.  Not surprisingly, our conclusion has to be that the radiative efficiency of the emission process responsible for the weak LHS X-ray flux is low.  The point here is that regardless of how well (or not, as the case may be) we understand the exact inner workings of the jets or accretion flows, we are probably not wildly off in terms of their dependencies and global scalings, which are relatively model independent as long as certain basic assumptions about conservation and power budgets hold.  Perhaps more importantly, these results demonstrate that understanding the accretion/ejection process in black holes, of all scales, irrefutably requires a multiwavelength perspective.  With single-band or non-simultaneous studies, we would never have discovered the correlation, nor would we have been able to derive such exacting constraints on accretion flow efficiency very close to the black hole.  It is interesting that in order to understand the regions closest to the black hole, we seem to require information about emission from regions which lie \emph{well beyond the gravitational reach of the black hole itself}.  Such a scenario would likely have not been envisioned even a decade ago, and the studies building on the discovery of the multiband correlations represent the significant evolution in our outlook over a relatively short period of time.  On the other hand, the premise behind AGN feedback's potential role in galaxy evolution necessitates such a link between the inner gravitational radii and the largest system scales, obviously in the form of the jets.  It would naturally be appealing if physical trends such as this correlation could be extended to the larger physical scales, in order to place better constraints on accretion processes in galactic nuclei.  

\subsection{Mass scaling and the fundamental plane}\label{subsec:fp}
\index{scaling}\index{fundamental plane}

Because most physical quantities in accretion models can in general be expressed via mass-scaling variables, it is possible to make predictions for the effect of mass on, e.g., the radio/X-ray correlation. It turns out that all black holes are predicted to follow a similar radio/X-ray correlation --- at least as long as they are in the equivalent of the LHS --- but the normalization of the relation will be inversely proportional to the mass.  The exact dependence can be predicted theoretically using scalings as above, and then tested against a sample of relevant (in this case, weakly accreting) AGN\index{AGN} such as LLAGN, and likely FR Is and BL Lacs.  This process was first carried out by two independent groups \cite{MerloniHeinzDiMatteo2003,FalckeKoerdingMarkoff2004}, and the relationship between radiative power, black hole mass and accretion rate was dubbed ``the fundamental plane of black hole accretion''.  More details about the FP are provided in Chap.~5, however there is quite a bit of scatter in the correlation for AGN even after ``mass-correction'' for comparison with BHBs, and it is difficult to separate out the possible effects of spin in AGN samples, which obviously does not play a role in correlations measured from individual sources.  Despite the complications the fundamental plane has been a rousing success for theory, and has opened up a new avenue for studying accretion because it allows us to compare behavior at very different size- and timescales.  We can essentially take the best of both worlds, studying for instance jet formation and evolution directly in the fast-evolving BHBs, while actually imaging regions close to the event horizon in nearby AGN.  

There are, however, important caveats to using data from AGN samples, not least of which are the lack of simultaneity and the lack of comparable spatial resolution. Even though the timescales are much longer than in BHBs, there is enough variability particularly for AGN at the smaller end of the mass range, that non-simultaneous data can skew estimates of the indices for the fundamental plane.  For instance LLAGN routinely display at least 20\% variability on a timescale of months \cite{Falckeetal2001,Nagaretal2002}, sometimes up to 50\% as seen in the UV cores of several AGN \cite{Maozetal2005}.  In \cite{Markoffetal2008} we found that repeating an earlier analysis of the mass-dependent coefficient in the fundamental plane, using simultaneous data instead of averages derived from the literature, resulted in a 50\% change in its value!  Combined with the different size scales leading to different predicted lags between the various bands of emission, it is quite difficult to be sure that we are always comparing apples to apples.  In a BHB we are likely averaging over waves of variability in the radio/IR emission of the jets, while in a LLAGN we can detect this motion explicitly over timescales of weeks to months.  Eventually with better data and statistics, we should be able to factor any lags into the analysis, but it is something to keep aware of as a factor in divergent values for correlation slopes, as well as other trends. 

For the brightest AGN with the largest scale jets, there is a more significant problem.  Many of the radio measurements are still taken with instruments like the VLA, which does not provide sufficient spatial resolution for sources residing at very high redshifts.   When we study a BHB, the jets are small enough (AU to pc) that achieving causality between the core and outer jets is never much of a concern; we can be assured that all jet activity is related to a recent outburst or flaring.  However the radio flux we measure from a distant AGN may include emission from plasma far out in the jets that is associated with activity in the nucleus thousands or more years ago, even from entirely different outburst cycles of accretion activity and reactivated by, e.g., internal shocks.  Accidentally integrating over several outbursts in bright radio-loud AGN could impact our attempts at associating them with BHB states.  For instance one of the major outstanding questions is why radio loud jets are present in some high accretion rate active nuclei, when this does not seem to be the case in BHBs except for a small period of transitional activity between the HIMS and SIMS.  There are two possible explanations, assuming one believes the states are shared in all black holes: either these are short-lived states (by AGN standards, still steady for us) that we are catching in the act, or we are indeed associating prior radio activity with a currently ``HSS-type'' nucleus that is no longer powering the jets.  Therefore it is important to push for VLBI imaging whenever possible for distant luminous radio galaxies used in comparisons with BHBs.

As a final note I think it is important to recognize the process by which the fundamental plane was discovered.  Interesting trends using new observational techniques were discovered, but a theoretical framework for interpreting and extending the results was necessary to fully exploit the data.  At the moment we are at something of an impasse in extending our mappings of particularly the transitional BHB states to AGN classes, because our theory has yet to catch up with the data.  The LHS/LLAGN (+FRI/BL Lac?) mapping was in many senses the easiest one, because we could assume a semi-steady state, and high optical depth.  In order to proceed further, I believe that we will first need theoretical models that can explain the transitional states, and that means understanding significantly more than we do about internal jet physics and launching processes.  
So while it is important to pursue better quality samples of AGN with good multiwavelength coverage, ultimately there is still hefty progress to be made on the accretion/ejection modeling.

\section{Modeling and hysteresis}\index{hysteresis}

\subsection{A brief overview of recent models}\label{subsec:models} 

So far I have not said much about modeling, other than to try to document how relatively simple models have been rather successful at helping us understand overall trends in the data.  Obviously there are a myriad of models at the moment for various aspects of accretion theory and only so many pages in this article, so I will focus on the ones which seek to explicitly address, in some quantitative way, the accretion/ejection connection.  Because the radio emission is generally agreed to originate in the jets, this bias naturally selects out that subset of models which attempts to link the X-ray producing processes directly to the outflows in order to reproduce observables such as the radio/X-ray correlations.   At the time of writing all such models are necessarily somewhat contrived, reflecting our rather limited basic understanding of the exact physical mechanisms linking the inflow and outflow.  In order to bridge this gap, any model must make various strategic assumptions for the physics of jet launching, in the hopes that their validity can be gauged by comparison with the data.  

The most ``fundamental'' approach is to attempt to model the plasma physics of the accretion process, often requiring the use of complex (M)HD codes, although significant progress can sometimes be made analytically.  Both techniques also involve approximations to the physics,  for simulations because otherwise they are too computationally expensive, and for more analytical calculations because otherwise they are not solvable without a simulation.  
In general these approaches focus usually more on the plasma dynamics rather than radiative physics, and are thus not optimized for direct comparison with spectral data.  Because this chapter focuses more on spectral calculations, I will only touch on some examples of these two approaches, which are discussed in detail in Chaps.~9 \& 10.   

Most models so far generally focus on some form of shared power budget as a way to link the jets and the accretion inflow, emphasizing uncovering the process that binds them and accounting for overall trends rather than explicit spectral/timing models.  However some classes of models start by assuming a specific energetic relationship, and then derive exact spectral predictions.  Having complementary approaches is always valuable, and so in the following I will briefly summarize the various modeling strategies currently underway, roughly grouped by conceptual framework.  

\vspace{.1in}\noindent\textbf{\textit{Jet launching scenarios}}\newline
After the discovery of jet quenching in BHBs during outbursts, as well as indications of a possible link between observed accretion states and recession of the thin accretion disk \cite{EsinMcClintockNarayan1997}, it seemed logical that the geometry of the accretion disk should be linked somehow to jet production.  Two papers earlier this decade explored, using rather simple and elegant arguments, how the disk geometry together with magnetic field configuration could affect the resultant jet power, and have strongly influenced the thinking of many subsequent groups.  In \cite{Meier2001}, Meier considered the case in which jet power is mostly dependent on the poloidal magnetic field strength, and to some extent also the field angular velocity in the inner regions, and explored the results for thin and thick disk geometries.  He argued that only geometrically thick flows such as RIAFs (or magnetic coronae, see below)\index{RIAF} could produce poloidal fields powerful enough to lead to the efficient radio-producing jets observed in, e.g., hard states.  A similar conclusion was derived by \cite{LivioPringleKing2003}, in the context of exploring accretion disk dissipation, and the variability of GRS~1915+105, and to this date most models rely on the interplay of accretion disk geometries to aid in explaining the launching scenario.

Several groups have in the meantime been working towards more detailed semi-analytical scenarios, in order to try to quantify more specifically not only the jet powers but also to understand the various BHB states and their relation to accretion disk physics.  One of these, \cite{VarniereTagger2002} based on earlier work by \cite{TaggerPellat1999}, suggest that an ``accretion-ejection instability'' in magnetized accretion disks can solve the problem of how disks transfer angular momentum vertically into jets rather than only outwards within the plane of the disk.  If the disk is assumed to be threaded by a poloidal magnetic field, then a spiral density wave instability can generate a Rossby vortex at the corotation radius.  The motion of the field footpoints due to the magnetic stress transports accretion energy and angular momentum from the inner disk to the corona via Alfv\'en waves.  If this corona feeds the jets, this mechanism can provide a link between the inflow and outflow.  It has also been proposed as a source of the QPOs\index{QPO} observed in BHBs \cite{Taggeretal2004,TaggerVarniere2006}.  While this is a very interesting scenario, it has yet to be verified to occur in 3D (i.e., simulations are not yet showing such instabilities).  Similarly it is difficult to compare to data as most of the conclusions are not yet predictive for generic sources.  However it is important to note that such instabilities likely must play a role in some of the explosive phenomena we observe, and the complex physics involved makes it challenging to calculate exact predictions.

\begin{figure*}
\centerline{\includegraphics[width=\textwidth]{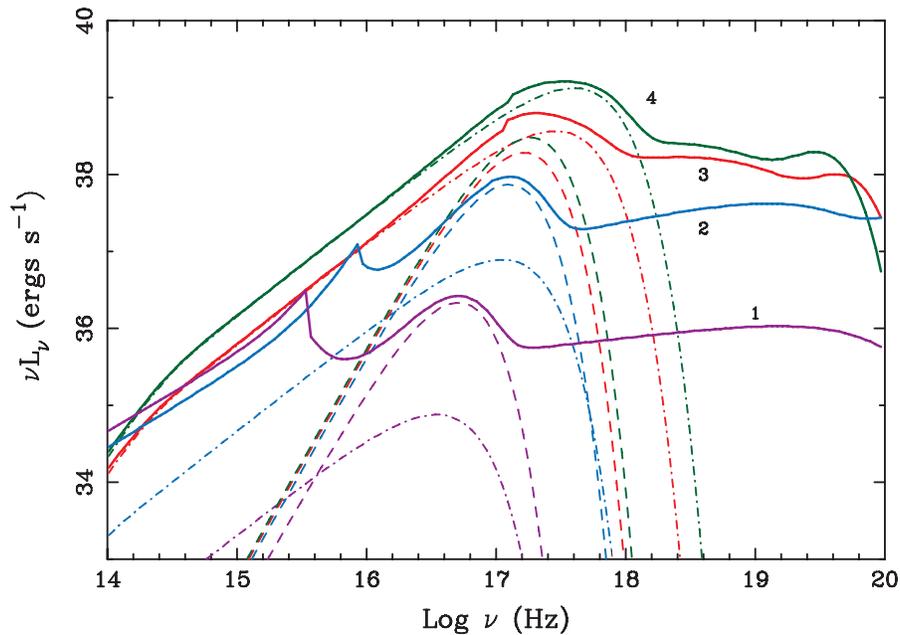}}
\caption{Predicted broadband SEDs for a thermally-driven jet combined with a magnetic corona model of a $10M_\odot$ black hole \cite{MerloniFabian2002}.   Solid lines 
show the total spectra, dashed lines show the contribution of thermal reprocessed radiation from the accretion disk and dot–dashed lines show the multicolor blackbody disk emission.  The accretion rate is increasing from 1 to 4 and colors (electronic only) indicate component/spectrum groupings.  The X-rays originate from the inverse-Compton upscattering of thermal disk and coronal synchrotron photons in the magnetic corona. The jet luminosity has been estimated assuming a flat spectrum up to the coronal self-absorption frequency and a radiative efficiency of 10\%.  The spiky peak at $10^{15}-10^{16}$ Hz in spectra 1 and 2 is the self-absorbed synchrotron emission from the magnetic corona itself, dominating the seed photons for Comptonization in spectrum 1 and still significant for spectrum 2. These SEDs do not include coronal bulk motion as suggested by \cite{Beloborodov1999}, which would significantly reduce the reprocessed emission (dashed lines) and thus extend the domination of synchrotron photons as Comptonization seed photons up to higher accretion rates. }
\label{fig:mf02}
\end{figure*}

Another line of approach is a magnetic corona, as mentioned in Sect. \ref{sec:paradigms}.  In the scenario explored by \cite{MerloniFabian2002}, a standard thin disk \cite{ShakuraSunyaev1973} at low accretion rates experiencing magneto-rotational instability (MRI) is ``sandwiched'' by a patchy magnetic corona, into which it dissipates a significant amount of its gravitational energy via buoyant flux tubes.  The rate at which energy is fed into the corona via dissipation increases with higher viscosity, and the speed at which magnetic structures are buoyantly transported upwards.  Both of these increase at lower accretion rates, and the dissipated energy will either heat the corona, or launch outflows.  Coronae with larger radii and scale-heights will result in more powerful outflows, and both will contribute to the total spectrum via synchrotron emission.  The authors present some sample spectra that already indicate the expected correlation between X-ray flux and hardness with optically thick jet emission (Fig.~\ref{fig:mf02}).  This baseline model was expanded upon in \cite{MalzacMerloniFabian2004}, where the consequences of this ``commmon reservoir'' of magnetic energy feeding both the corona and jets were applied to explain the puzzling optical/UV/X-ray variability in XTE~J1118+480\index{XTE~J1118+480}.  In their proposed scenario, the corona produces the X-ray via standard Comptonization of lower energy seed photons from the geometrically thin disk, while the jets produce the optical via synchrotron radiation.  In this way, the lag between the optical and X-ray is explained by the displacement of plasma outwards from the corona into the jets, while energetically they share a common input of power from the magnetic processes generated in the thin disk.

Finally there is the series of works building on
\cite{FerreiraPelletier1993,FerreiraPelletier1995,Ferreira1997},
culminating most recently in \cite{Ferreiraetal2006}, which describes
a complicated phenomenology to tackle the various states observed in
BHBs.  Within a standard accretion disk, they posit that a magnetized region forms
where the jet is launched, called the ``jet-emitting disk'', or JED.
The inner part of the jet is a pure pair outflow that is accelerated
to high bulk Lorentz factors, surrounded and confined by a slower,
matter-dominated jet, itself self-confining.  This scenario bears some
similarity to the ``magnetically-dominated accretion flow'' (MDAF)\index{MDAF}
solution described in \cite{Meier2005} and explored for the first time
numerically in 2D using a fully general relativistic MHD code
including radiative cooling (COSMOS++;
\cite{AnninosFragileSalmonson2005}) by \cite{FragileMeier2009}.  It
is interesting to note that if radiative cooling is self-consistently
included, magnetic pressure dominates the gas pressure near the inner
regions of the inflow, and synchrotron cooling becomes very important
for the dynamics.  The advantage of the simulations is that they can study jet formation directly from the given initial conditions, but linking them to reality can be challenging.  Often it can be too computationally expensive to run the simulations long enough to see a steady state form.  My colleagues and I are currently developing several new methods, both semi-analytical and using simulations, to try to address this gap.

\vspace{.1in}\noindent\textbf{\textit{Jet-disk symbiosis }}\newline
In the meantime, there is always a need to understand the boundary conditions and environment in order to guide the necessary assumptions in the above classes of approach.  From this more phenomenological point of view, aimed at fitting observations, some of the earliest quantitative work on the self-coined ``jet-disk symbiosis'' can be found in \cite{FalckeBiermann1995,FalckeBiermann1999}.  By assuming that the jet power is linearly proportional to the accretion rate in the inner disk, and assuming a freely expanding jet with maximal efficiency, they were able to extend the earlier work on very idealized jets by \cite{BlandfordKoenigl1979} by solving for the velocity profile and including the resultant cooling terms.  The lack of knowledge about the exact launching geometry and mechanism is channelled into several free parameters determining the jet initial conditions, but the exact nature of the jet/disk link cannot be derived from this model because the disk power mainly serves as an ``accretion power reservoir'' for the jets. Although mostly applied to AGN, this work provided the groundwork theory for the scalings used to determine the Fundamental Plane described above.   

\begin{figure*}
\centerline{\includegraphics[width=0.75\textwidth]{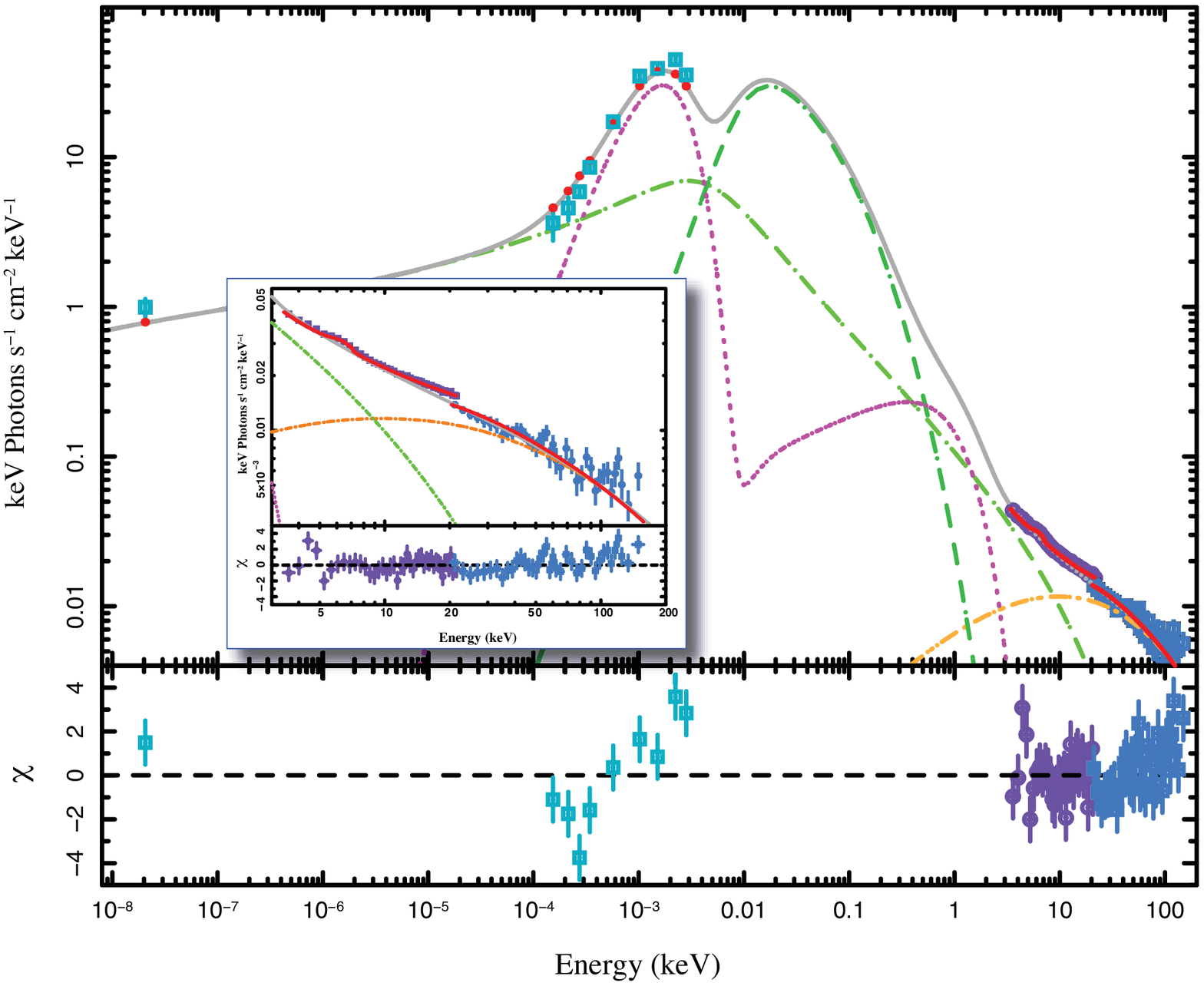}}
\centerline{\includegraphics[width=0.8\textwidth]{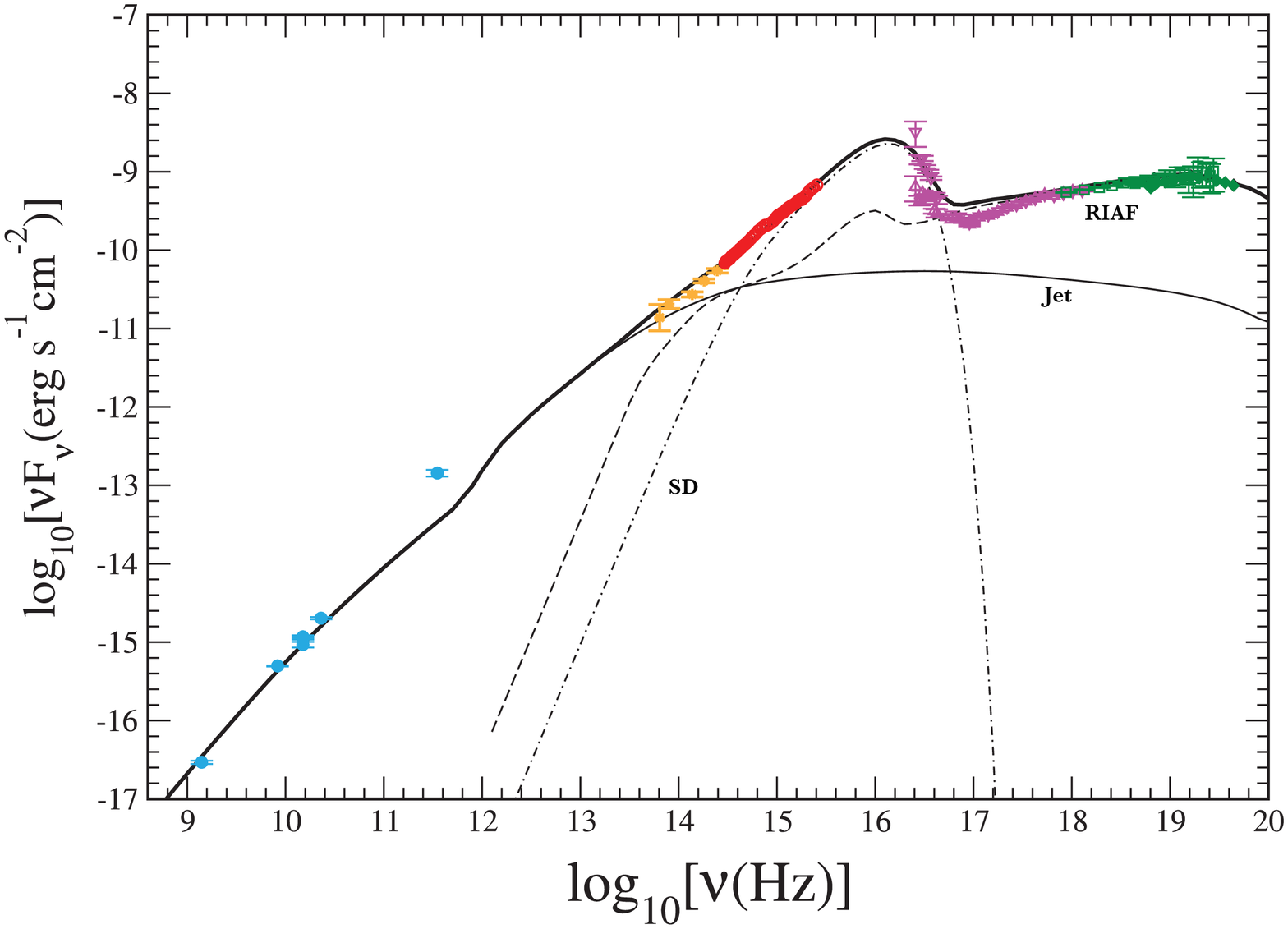}}
\caption{\emph{Top:} Outflow-dominated model \cite{MarkoffNowakWilms2005} fit to the simultaneous broadband data from a hard state observation of GRO~J1655-40\index{GRO~J1655-40} taken on 24 Sep 2005, with the VLA, Spitzer, SMARTs and RXTE.  The dark (green in electronic version) dashed line (labelled PrS in print version) is the pre-shock synchrotron component, the lighter (green) dash-dotted line (labelled PoS in print version) is the post-shock synchrotron component, the (orange) dash-dotted line (labelled Com in print version) represents the SSC plus the disk external Compton component, the (magenta) dotted 
line (with peaks labelled BB and Str in print version) is the multi-temperature disk black body plus a black body representing the companion star in the binary system. The solid (red) line/points indicate the total model after including absorption and convolution with reflection and the RXTE response matrices, while the individual components and (grey) total spectrum do not take these into account.  See \cite{Migliarietal2007} for more details.   \emph{Bottom:} A joined accretion disk-jet model applied to the broadband data from XTE~J1118+480\index{XTE~J1118+480} (color in electronic version). The dashed (labelled RIAF) and dot-dashed (labelled SD) lines show the emission from the hot (RIAF) and cool accretion flows, respectively. The thin solid line shows the emission from the jet, and the thick solid line indicates the total spectrum.  For details see \cite{YuanCuiNarayan2005}.}
\label{fig:specmods}
\end{figure*}

The model explored for BHBs in \cite{MarkoffFalckeFender2001}, and the first attempt at explaining the radio/X-ray correlation \cite{Markoffetal2003}, were based on these same general assumptions but included more detailed internal physics and radiative processes.  One of the most important conclusions of this work, as mentioned in Sect.~\ref{subsec:jetspec}, is that for reasonable energy inputs the compactness of BHBs compared to AGN even at low luminosities leads to a prediction of significant synchrotron X-ray emission from the jets in BHBs.  At the same time, the presence of reflection features observed in BHB hard states (e.g., \cite{ZdziarskiLubinskiSmith1999}) argues that a significant portion of the hard X-rays cannot be too strongly beamed, or originate too distant from the disk.  The weakness of the reflection in the LHS, however, does imply that the covering fraction cannot be too high. \cite{MarkoffNowak2004} explored this issue and determined that to reproduce the observed reflection fractions in the context of an jet-dominated model, the base of the jets would need to be compact enough to produce high-energy photons via Comptonization of either internal synchrotron photons, or external disk photons.  This idea was explored in more detail in \cite{MarkoffNowakWilms2005}, where we showed that this base region can in fact mimic the observational features attributed to an accretion disk corona, however with different assumptions than those required for thermal disk Comptonization models \cite[e.g][]{HaardtMaraschi1991}.  Specifically, because of the relativistic flow in a direction away from the disk, the photon field for upscattering is quickly dominated by synchrotron photons, and thus the thermal disk spectrum is hard to constrain via its weak spectral signature alone.  On the other hand, the strength of this model is how it demonstrates that a magnetized corona or RIAF-like\index{RIAF} flow at small radii can be consistent with directly feeding the jets, while also providing a very good description of the radio through X-ray data (see also \cite{Galloetal2007,Migliarietal2007,Maitraetal2009}.  The soft X-rays contain a significant synchrotron radiation component, becoming increasingly inverse Compton-dominated towards the hard X-rays.  Thus one prediction of this model is that the exact slope of the radio/X-ray correlation will vary depending on which component dominates the X-ray band under consideration.  While synchrotron predicts a slope of 0.7, synchrotron self-Compton SSC will produce a slope closer to 0.5, thus combinations can give intermediate values, and this may contribute to some of the variations in slope recently reported (Corbel et al. 2008, in prep.).  
The limitation of this model is that, by assuming the disk/jet linkage \textsl{a priori}, a detailed understanding of their relationship cannot be tested beyond a consistency check.  Similarly, the fact that the model assumes a steady-state precludes it from being tested against timing data such as frequency-dependent lags.  Another steady-state approach to fitting broadband SEDs can be found in \cite{YuanCuiNarayan2005} and \cite{YuanCui2005}, who have developed new models based on earlier work \cite{YuanMarkoffFalcke2002,Yuanetal2002}, that pair a simple jet model with a more detailed treatment of a RIAF.  In this case, however, the Comptonizing region belongs predominantly to the accretion inflow.  The strength of both of these model classes is their ability to test basic assumptions about the link between inflow and outflow components against the high-quality broadband data now available (Fig.~\ref{fig:specmods}).

\vspace{.1in}\noindent\textbf{\textit{Other modeling approaches}}\newline
Several other groups have honed in on the modeling of more specific behaviors.  For instance \cite[e.g.][]{ChattopadhyayDasChakrabarti2004} have an almost inverted approach to those described above, in that they begin by assuming the presence of a pair jet at the inner accretion flow, and explore the energetics and mechanisms involved in accelerating and confining it via radiation pressure from the accretion flow.  The authors focus thus in detail only on the accretion flow rather than the jets, in the context of a two component advective flow (TCAF; \cite{Chakrabarti1996,ChakrabartiTitarchuk1995}), resulting in the formation of what they term a centrifugal boundary layer (CENBOL)\index{CENBOL}, that acts as the ``corona'' for producing the hard X-rays via Comptonization.  This class of models has been invoked to explain the observed pivoting in the X-rays that appears to be correlated with the radio flux in Cyg~X-3\index{Cyg~X-3} and GRS~1915+105\index{GRS~1915+105} \cite{Choudhuryetal2003}.  The physics and launching of the jets is not addressed explicitly.

One of the more radical proposals comes from \cite{RobertsonLeiter2002}, who suggested that any jet-producing object (driven by magnetic fields) could be interpreted instead as a ``magnetospheric eternally collapsing object'', or MECO, as long as the magnetic moment is sufficiently high. For sources we interpret as BHBs, they predict a field strength around $10^8$ G, quite a bit higher than derived from standard equipartition arguments in plasma, but not inconsistent with the assumed energy budget.  The idea is a new application of magnetic propeller theory, where the inner disk is coupled to the interior magnetosphere, and in \cite{RobertsonLeiter2004} the same authors actually derive the scaling laws for the radio/X-ray correlation and dependence on mass for their favored scenario.  They show that for certain assumptions, they can indeed reproduce the relations determined empirically as discussed above, and in Chaps.~4 \& 5.  Detailed spectral fits have, however, thus far not been presented, nor is it clear that this model can survive frequency-dependent size constraints such as those now possible with Sgr~A* \index{Sgr~A*}\cite{Boweretal2004,Shenetal2005,Boweretal2006,Doelemanetal2008}.  

Another model that focuses more on explaining the timing features of BHBs than detailed spectra is that of \cite{GianniosKylafisPsaltis2004}.  Taking a simplistic formulation for the jet, they obtain results that are consistent with both the soft-hard X-ray lags as well as energy-dependent trends in the autocorrelation function \cite{MaccaroneCoppiPoutanen2000}, while also reproducing trends in the evolving X-ray spectra.  In a followup paper, \cite{Giannios2005} consider the effects of a power-law distribution of electrons in the jets, and demonstrate that this model can also account for the multi-wavelength features in BHB spectra.  However, it is not clear if this particular choice of configuration can successfully reproduce the radio/X-ray correlations.  

And most recently, another range of jet models have been developed with the aim especially to tackle the issue of whether or not microquasars can produce $\gamma$-rays, as some recent papers have claimed \cite[e.g.][]{Paradesetal2000,Albertetal2007}.  While high-energy protons are considered explicitly as sources of high-energy secondary particles, via their collisions, in \cite{Romeroetal2005} (as compared to the Falcke \& Biermann, Markoff et al. papers, where protons are included only as kinetic energy bearers), the focus was not on the broadband spectrum or the link to the accretion flow.  However recent work by \cite{Bosch-Ramonetal2005}, and several subsequent papers based on this model,  explores several potential scenarios for high-energy photon production, where the jet power is assumed proportional to the inflowing accretion rate.  In general they consider cases with a sub-equipartition magnetic field, that may influence their findings in favor of external photons as sources of Comptonized flux, but it is interesting nonetheless, especially with the recent launch of the \textsl{Fermi} observatory.  With new broadband campaigns being planned that include \textsl{Fermi} as well as the ground based TeV instruments H.E.S.S. and MAGIC (and eventually the CTA), along with particle detectors like Auger, ANTARES and IceCube, it is extremely important to understand whether BHBs can be significant sources of cosmic rays and neutrinos.
 
\subsection{Hysteresis}\label{subsec:hyster}\index{hysteresis}

\begin{figure*}
\centerline{\includegraphics[width=0.85\textwidth]{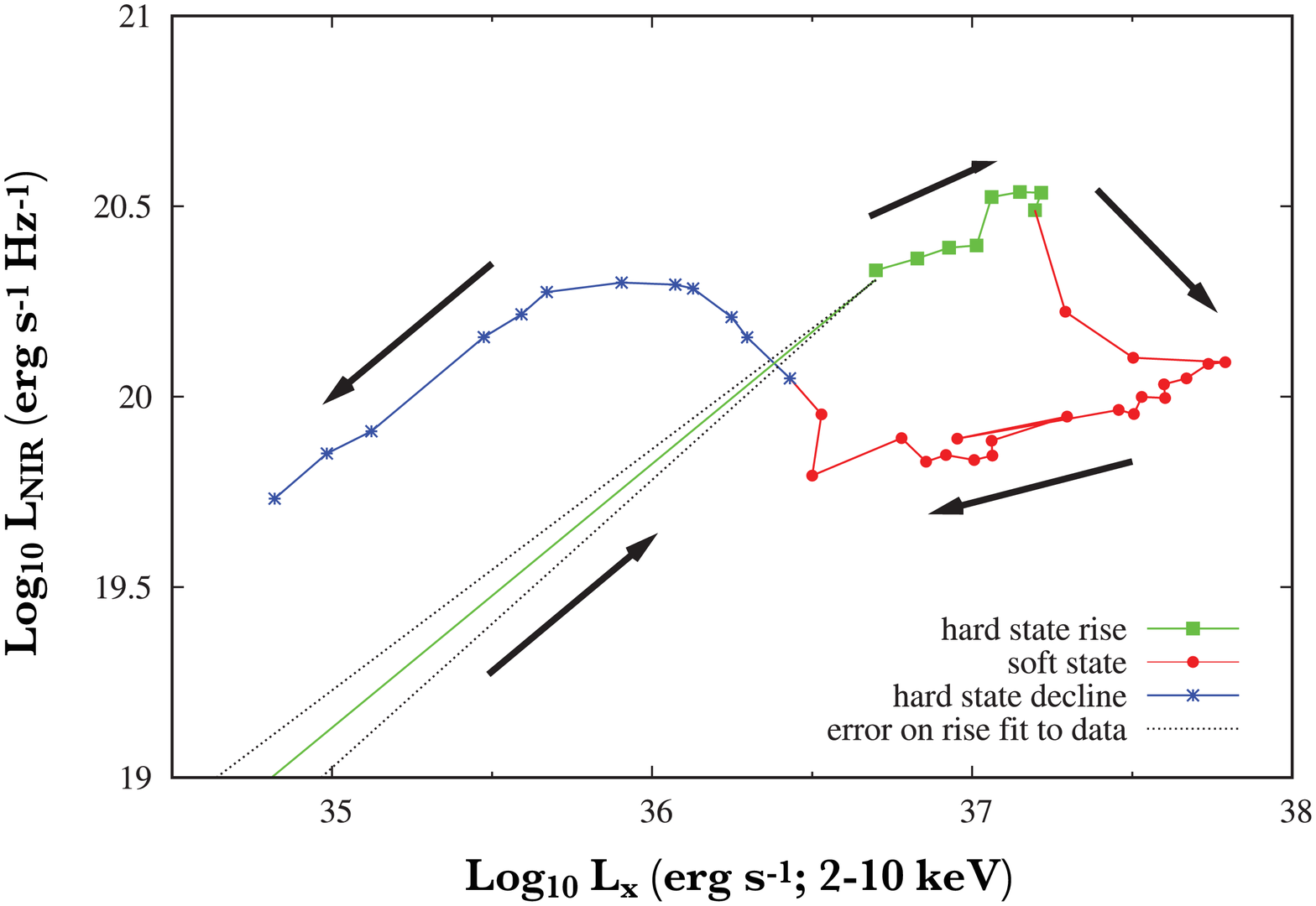}}
\centerline{\includegraphics[width=0.85\textwidth]{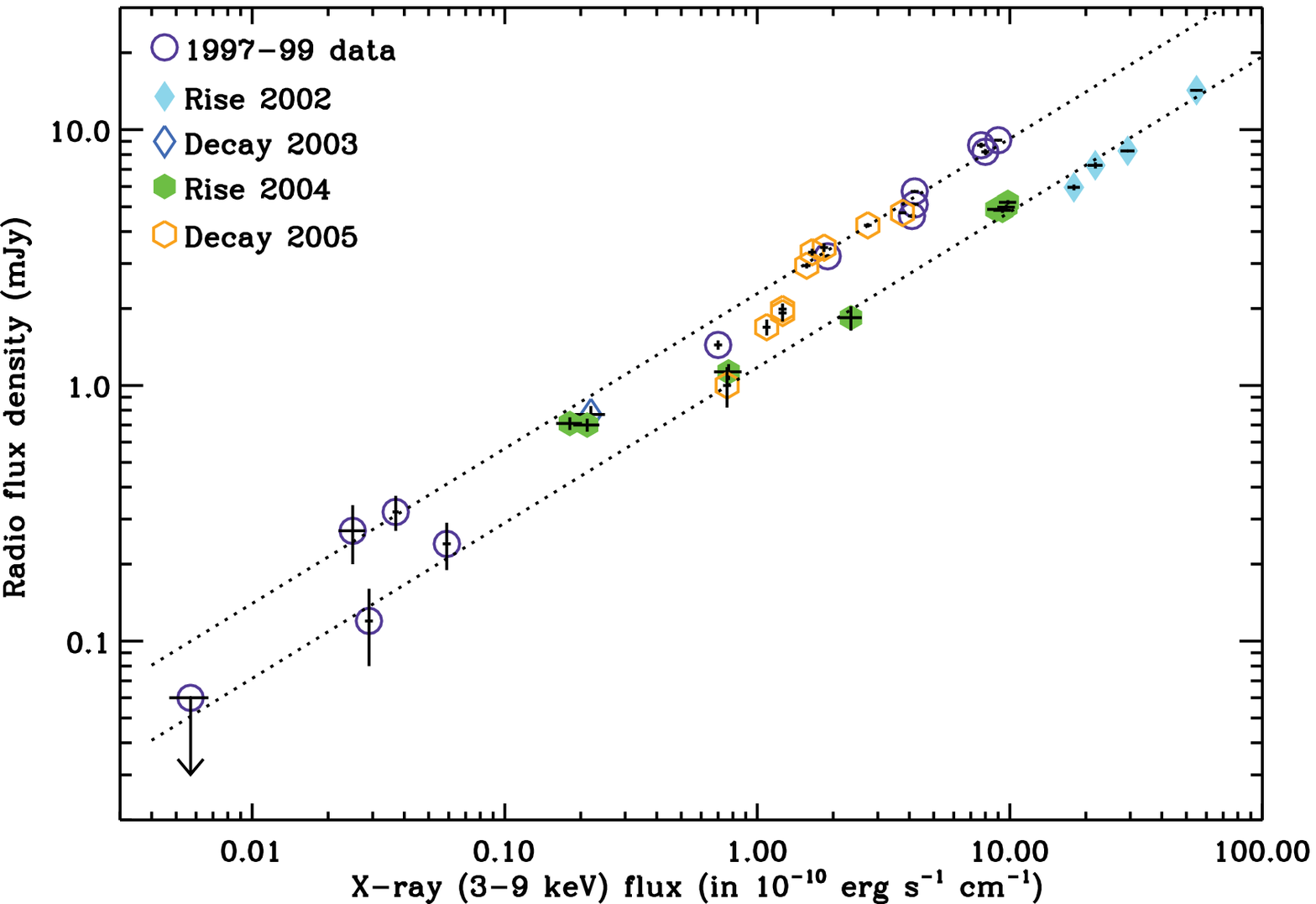}}
\caption{\emph{Top:} The hysteresis in the infrared vs. X-ray
  quasi-simultaneous luminosities of the BHB XTE~J1550-564\index{XTE~J1550-564} during its
  outburst of 2000. In at least bright hard states, the (N)IR emission is dominated by
  the jet. The jet is quenched in the soft state so the infrared
  drops, and is then dominated by light from the accretion disk. At a
  given X-ray luminosity, the infrared emission appears to be weaker
  in the rise of the outburst compared with the decline. The power law
  indicating the slope of the rise correlation is derived from fits to
  the X-ray and infrared exponential rise light curves. From
  \cite{Russelletal2007}. \emph{Bottom:} Hysteresis in the radio
  vs. X-ray simultaneous fluxes of the BHB GX~339-4\index{GX~339-4}.  The rise
  phase always seems to follow a lower luminosity track for
the same X-ray flux compared to the decay phase (from Corbel, Coriat
et al., in prep.).}
\label{fig:hyster}
\end{figure*}

As discussed elsewhere in this volume, the hysteresis observed during
the typical BHB outburst cycle, as evidenced especially by the ``q'' shape in the hardness-intensity diagram, signals that at least some other factor besides the accretion rate is playing a major role in the source evolution.  One strong candidate would be magnetic field configurations in the accretion disk, but this question is by no means settled.  Interestingly, hysteresis has also recently been discovered in the radio/X-ray and IR/X-ray correlations found in the hard state during outbursts; some examples of this are shown in Fig.~\ref{fig:hyster}.  It is now clear that the system does not return during decay to the same hard state configuration that it had during the rise stage.

Fig.~\ref{fig:hyster} shows two examples of hysteresis, in the radio/X-ray and IR/X-ray correlations for two BHBs.  The difference in IR flux between the hard and soft state could be interpreted as due to the quenching of the compact jet synchrotron emission during the hard-to-soft transition.  
Within the hard states themselves, however, the IR flux is clearly higher during the decay compared to the rise in outburst, a trend that is also seen in the radio/X-ray correlation hysteresis.  In general the slope of the correlation appears to steepen below around $10^{-4}\;{\rm L}_{\rm Edd}$, at least during the decay phase, but these results are new and such hysteresis has not been measured for many sources at this point.  It is already clear that the variation in correlation slope as the system ``jumps'' tracks of slope $\sim0.6-0.7$, as well as the difference in normalization for the two tracks, will have implications for the Fundamental Plane\index{fundamental plane} if AGN experience something similar.  Perhaps some of the scatter observed in that correlation could be due to a similar effect.  But perhaps more importantly, this hysteresis must provide information about the second physical parameter driving the outburst in addition to the accretion rate.  Furthermore there is preliminary evidence that for some source outbursts, the hysteresis is not seen at all in the IR/X-ray correlation, although present in the radio (Corbel et al., in prep.).  

As discussed in Section~\ref{subsec:interp}, changes in the correlation slope could be rooted in a change in the efficiency of the radiative process.  One possibility could be that the X-ray band contains emission from two different components, that are swapping dominance during the decay.  In their analysis of $\sim200$ pointed \textsl{RXTE} observations of Cyg X-1, \cite{Wilmsetal2006} found the best fits with a broken power-law where the indices of the two components were strongly correlated.  In my colleagues' and my interpretation, such a correlation would be an indication of correlated synchrotron and inverse Compton radiation, where synchrotron photons dominate the Compton photon pool.  A flattening in the slope of the radio/X-ray correlation at low luminosities (i.e., from 0.7 to 0.5) could for instance indicate that SSC cooling was the preferred channel rather than synchrotron cooling.  But this explanation does not account for the hysteresis in correlation normalization itself.  Most likely the change in normalization relates to physical differences between the inner disk/corona and jets during the rise and decay, such as different magnetic field configurations perhaps resulting from changes in the inner disk structure.  At the time of writing we do not yet have an explanation for this phenomenology, but interestingly this trend may be consistent with some other phenomena observed at the lowest luminosities, as discussed briefly below.

\subsection{Is quiescence distinct from the hard state?  Sgr~A* vs. A~0620-00}
\index{Sgr~A*}\index{A~0620-00}

As a final point, I would like to mention some additional aspects of jet physics where we might glean some insight by studying two sources at the very bottom of the accretion range, Sgr A* and A~0620-00.  By virtue of the FP, we can also compare these black holes to each other in order to derive a better picture of the buildup of accretion activity at the lowest levels.  

The change in the correlation towards quiescent luminosities, as well as reports of X-ray spectral steepening \cite{CorbelTomsickKaaret2006,CorbelKoerdingKaaret2008} has led to some speculation that quiescence is not simply a smooth continuation of the hard state down to lower accretion rates.  On the one hand, it seems that steady radio jets continue to exist down to at least X-ray luminosities of $\sim10^{-8.5}$ L$_{\rm Edd}$ \cite{Galloetal2006} and that the correlation is maintained to quiescent level accretion rates, at least for the one source for which there are good data.  The persistence of the same correlation down into quiescence would argue against scenarios such as \cite{YuanCui2005} mentioned above, whose coupled RIAF-jet model predicts a turnover to a form $L_R\propto L_X^{1.23}$ somewhere below $10^{-5}-10^{-6}$ L$_{\rm Edd}$.  On the other hand in some sources we do see the radio begin to drop faster compared to the X-ray emission at higher accretion rates than classically considered quiescent, though not exactly with the same 1.23 index.  However, the data are currently not sufficient to determine if there is also a mass dependence of this turnover, which could explain some of these differences from source to source.

\begin{figure*}
\centerline{\includegraphics[width=.8\textwidth]{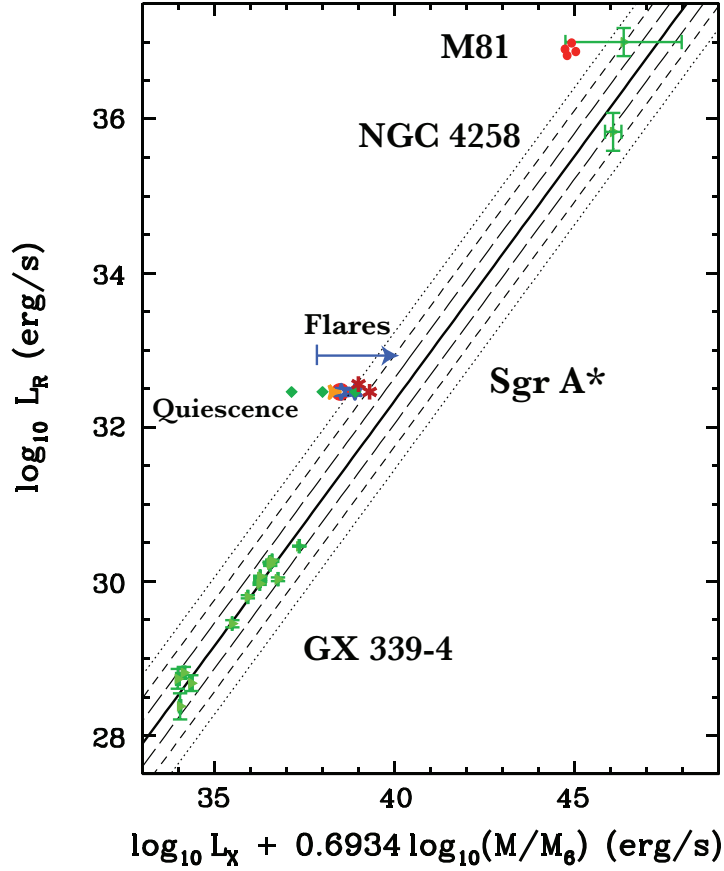}}
\caption{ The fundamental plane (FP) radio/X-ray luminosity correlation for the three best-measured
  sources bracketing (in scaled flux) the Galactic center supermassive black hole, Sgr
  A*: the BHB GX~339-4, and the LLAGN NGC~4358\index{NGC~4358} and M~81\index{M~81}.  Both
  quiescent and flaring states of Sgr~A* are indicated.  The 
 two LLAGN data points and their respective error bars represent the average and rms
  variation of all prior non-simultaneous observations.  For M~81, the
  four dark (red in electronic version) points indicate results from simultaneous radio/X-ray
  observations \cite{Markoffetal2008}.  The solid line indicates
  the best fit correlation using linear regression from Monte Carlo
  simulations of the data, with contours in average scatter $<\sigma>$
  from the correlation represented as increasingly finer dashed
  lines (figure modified from \cite{Markoff2005}).}
\label{fig:sgra}
\end{figure*}

The rise phase out of quiescence can also be very steep in the other direction: in several outbursts, the X-rays are seen to rise very rapidly while almost no change is observed in the radio band.  Such an extreme X-ray rise is especially interesting in the context of similar behavior in Sgr A*.  Sgr A* is the only accreting black hole, for which we have good radio and X-ray data, that does not fit on the FP\index{fundamental plane}.  Instead it sits in quiescence several orders of magnitude below the mass-scaled FP correlation, in the direction of weak X-ray flux.  However, during the approximately daily X-ray flares \cite[e.g.][]{Baganoffetal2001,Baganoffetal2003}, Sgr A* approaches the X-ray flux at which the FP radio/X-ray correlation would be predicted to take hold, though it has yet to come within an order of magnitude of actually achieving the emissivity required to test this idea (see, e.g., \cite{Markoff2005} and Fig.~\ref{fig:sgra}).  Even so, the sharp X-ray rise seems suggestively analogous to that displayed by the 2004 rising data for GX~339-4, shown in the bottom panel of Fig.~\ref{fig:hyster}.  If Sgr A*'s behavior is indeed a reflection of the same physics occurring in BHBs, then we could be witnessing its flickering attempts at outburst activity from out of quiescence, but that it is never sufficiently fueled to reach the sustainable levels of the FP.  Although we do not yet have an understanding of the mass-dependence of any steepening of the FP correlation, it seems likely that Sgr A* is sitting far below (in X-ray flux) the observations of the rise data from a BHB, when scaled to a comparable mass.  
Unfortunately equivalent luminosities ($L_X\leq10^{-10}L_{\rm Edd}$) would be difficult to observe for BHBs, and in observing long enough to obtain a significant measure of such an X-ray flux we would invariably be integrating well over any similar fast flaring timescales, when scaled for the smaller dynamical times.  Potentially the only way to test this hypothesis is to wait (and hope we observe) a flare in Sgr A* at least an order of magnitude brighter than any seen so far, assumedly quite a rare event if one assumes a power-law distribution in flare magnitude.  Perhaps we will get lucky and one of the S-stars will travel a bit too close during periastron, suffering tidal disruption and triggering a major accretion event (e.g., \cite{AlexanderLivio2001})!

\vspace{.1in}\noindent\textbf{\textit{Jet launching physics: clues from the lowest luminosities }}\newline
Aside from this speculation, Sgr A* is very interesting for understanding jet formation at the lowest accretion rates because it is so close that we actually have very high quality data, despite its weak (in Eddington units) emission relative to any other black hole we know.  Its radio through IR emission is clearly due to synchrotron radiation, given the polarization and high brightness temperature (see \cite{MeliaFalcke2001}, and references therein), and the presence of the signature flat/inverted radio spectrum likely indicates a weak, collimated jet outflow.  Such a jet has not yet been imaged, however, leading to suggestions that the radio emission originates in an uncollimated disk wind, despite many difficulties with this interpretation.  The opaque and stratified geometry required to reproduce a flat/inverted spectrum can in principle be fulfilled by a RIAF\index{RIAF} with synchrotron-emitting thermal particles, but the radial profile of such a flow would normally result in a steep falloff in the radio flux with decreasing frequency (see, for example, \cite{Narayanetal1998}).  The only way to avoid this problem is to posit the existence of an additional non-thermal power law tail in the electron distribution, that is somehow engendered in the thermal accretion flow \cite{YuanQuataertNarayan2003}.  To be fair, the amount of electrons required in the tail is very small, and thus minor magnetic events in the flow could potentially account for such limited particle acceleration.  However, this idea still seems rather ad-hoc given that it has only been necessary to invoke for the one source where any jets would be difficult to resolve through scattering effects in the Galactic plane.  Jets naturally provide a solution to the problem, as they are known particle accelerators, and in radio spectral shape and polarization Sgr~A* resembles another LLAGN in a similar grand design spiral galaxy, M~81*, where weak jets have been resolved \cite{BietenholzBartelRupen2004,Markoffetal2008}.   

The physics necessary to explain the lack of extended jets in Sgr A* also fits in with our understanding of particle acceleration near the black hole.  
It is interesting to note that Sgr A* also provides the most stringent limits on its internal particle distributions than any other black hole we can currently study.  While optical depth effects allow us to derive important scaling relations despite our lack of precise knowledge of such particle distributions in most sources, it hides details of the internal physics that would allow us to constrain particle acceleration or derive the total jet internal power due to lack of information about the minimum energy in the radiating lepton distribution.  In Sgr A* we can deduce directly from measurements that the particles responsible for the highest energy emission are predominantly thermal in nature, because any significant accelerated power-law tail would reveal itself prominently in the IR band.  In contrast, the observed IR spectrum reflects an extremely steep drop more consistent with weak to absent acceleration \cite{Genzeletal2003,Ghezetal2004}.  Thanks to advances in adaptive optics, along with identifying Sgr A* for the first time in the IR, these observations also led to the discovery of sporadic flaring events with marginal evidence for hardening in a power-law tail in the underlying particle distribution.  However statistics are still a challenge, and the two main observational groups do not yet agree on the exact spectral index during quiescence and flares.  

I propose that the flaring in both IR and X-ray is an indication that while plasma jets are still formed at low luminosities, the internal structures necessary to accelerate particles efficiently within them are not yet able to maintain themselves.  The resulting lack of a stable high-energy particle population compared to accreting black holes at higher accretion rates leads directly to a prediction of very compact jet photospheres.  Going back to Fig.~\ref{fig:jet}, the measured size of the photosphere at a given radio observing frequency is dependent on the broadband extent of the optically thin emission, in accreting black holes represented by power-law emission.  Therefore if the jets in Sgr A* contain predominantly thermal distributions of particles, the lack of a strong power-law tail would predict a more compact photosphere, and result in a less elongated Gaussian, difficult to distinguish from predictions of a RIAF.  Even small elongation would be easily hidden behind the scattering screen of intervening Galactic Plane material (see discussion in \cite{MarkoffBowerFalcke2007}).   One interesting test for the future will be to look for elongation in the photosphere measured with VLBI during bright X-ray flares, ideally at high frequencies such as reported in \cite{Doelemanetal2008}.

As for the flares themselves, they are providing hints that nonthermal processes also contribute significantly to the X-ray emission in both weak LLAGN as well as hard state BHBs.  Because of the lack of a thermal accretion disk as a source of thermal seed photons in Sgr A*, the most likely explanation for the simultaneous nonthermal IR/X-ray flares is the synchrotron self-Compton process \cite{Markoffetal2001,YuanQuataertNarayan2004}, although a direct synchrotron contribution may be present for very short episodes.  It is notable that all ``flavors'' of theory currently agree on this point, regardless of whether they favor inflow or outflow accretion solutions.  The submm/IR ``bump'' and non-thermal X-ray flaring emission must originate in a region of hot, magnetized plasma very close to the supermassive black hole, because of the short variability timescales (e.g., \cite{Baganoffetal2001}).  If the same geometry is present in BHBs at low accretion rates, as seems likely given the success of various scaling arguments, then a synchrotron/SSC contribution to the X-rays is even more likely because of the more compact relative scales. 

By looking at actual similarly weakly accreting BHBs when possible, we can attempt to test this theory.   As detailed in Section~\ref{subsec:models}, the same physical model for brighter weakly accreting states seems to well describe data across the mass scale.  To really probe quiescence is much more challenging in BHBs.  For instance, A~0620-00 is only several orders of magnitude brighter (in Eddington units) than Sgr A* in the X-ray band, but for a Galactic BHB this power translates into a very low count rate.  In \cite{Galloetal2007} we reported on preliminary results attempting to fit the rather poor quality X-ray spectrum, where we did find that an SSC-dominated model was favored, similar to the case in Sgr A*.  We also found evidence for weaker particle acceleration in comparison with other BHBs at higher accretion rates, supporting the interpretation that the processes responsible for particle acceleration in the jets breaks down below some critical $\dot m_c$, and which may bear some relevance for the behavior rising out of and decaying back into quiescence.  If acceleration is indeed associated with internal shocks, perhaps below $\dot m_c$ the flow does not have the initial conditions required for the necessary multiple-velocity ejecta.  Or it may be that the jet plasma simply can no longer be accelerated to bulk velocities beyond the fast magnetosonic point where shock-like disruptions could form.  In any event, it is an intriguing possibility that somewhere between A~0620-00 and Sgr A*, or between $L_X=10^{-10}-10^{-7}L_{\rm Edd}$, the particle acceleration process in the jets begins to break down.  With new all-sky monitors such as LOFAR soon discovering many new transients, we can hope to find an even more massive, closer BHB in quiescence to test some of these premises, at the same time that we are currently investigating some of the more theoretical aspects (Polko, Markoff, Meier, et al., in prep.).  

\section{Conclusions}\label{sec:concl}

As is often the case with theoretical studies, there is no definitive conclusion to this Chapter, other than to say that the story continues.  The last decade has been extremely productive in terms of furthering our understanding of the systematic behavior of BHBs, and in particular simultaneous multiwavelength campaigns have for the first time allowed us to connect accretion behavior over the extreme ranges of the black hole mass scale.  While the fundamental plane of black hole accretion provides a powerful tool for comparing the fast-varying trends in BHBs to the protracted duty cycles of AGN, it represents just the tip of the iceberg in terms of understanding the subtleties of potential mass and power scalings.  First of all, we now know that the underlying correlation for BHBs does not have exactly the same slope for all sources, though its spread is still rather narrow, and we see different normalization ``tracks'' during rise and outburst, that has yet to be factored into the AGN results.  But perhaps more importantly, when comparing BHB cycles based on observing individual sources to populations of AGN selected from samples, there are enormous elephants in the room such as how to account for the role of spin and the kinds of differences in accretion phases expected in AGN of various galaxy morphologies that should not be relevant for BHBs.  In order to assess the extent to which mass scaling translates into an exact AGN equivalent for every BHB state, we need to move away from the comfortable lower right hand side of the hardness-intensity diagram, away from steady state jets and towards a better understanding of the more extreme transitional behavior that occurs between the hard and soft accretion states.  Several groups are now currently tackling this progression, and armed with the experience and knowledge described here and elsewhere in this volume, I anticipate that we will be able to address these open issues within the next decade.  At the same time, it is important to note that the impending loss of the RXTE ASM with no immediate replacement will be an undeniable detriment to our ability to obtain the high-quality multiwavelength monitoring data that enabled these discoveries in the first place.  On the other hand, the all sky monitoring to find new classes of sources and transient behavior, now with $\gamma$-rays using Fermi as well as ground-based radio facilities like LOFAR and other SKA pathfinders, will undoubtably flesh out the picture we have now of black hole accretion in transition.  But aside from the high-quality multiwavelength data, I would like to end by returning once again to emphasize the theoretical framework that was necessary before the observed trends could be ``translated'' into a more global picture.  We need a new foothold on the theory of low-to-high accretion transition pathway, and once we have it this foothold will hopefully help open the door to our understanding not only of jet quenching in BHBs but also cycles of AGN activity producing super-cavities in clusters and potentially feedback.  I think that the small scales still have quite a bit left to teach us about the largest, and that we have a long but interesting way to go before we can consider our paradigms complete.

\bibliographystyle{abbrv}
\bibliography{chap06_markoff_refs}



\end{document}